\documentstyle[12pt,epsf]{article}
\newcommand{\HH}{{\cal H}}

\newcommand{\be}{\begin{equation}}
\newcommand{\ee}{\end{equation}}
\newcommand{\ben}{\begin{eqnarray}\displaystyle}
\newcommand{\een}{\end{eqnarray}}
\newcommand{\refb}[1]{(\ref{#1})}
\newcommand{\p}{\partial}
\newcommand{\sectiono}[1]{\section{#1}\setcounter{equation}{0}}

\begin{document}

\begin{flushright}
hep-th/0010240\\
MRI-P-001003
\end{flushright}

\vskip 3.5cm

\begin{center}
{\Large \bf Fundamental Strings in Open String Theory 

\medskip

at the Tachyonic
Vacuum}\\

\vspace*{6.0ex}

{\large \rm Ashoke Sen\footnote{{\tt ashoke.sen@cern.ch}, 
{\tt sen@mri.ernet.in}}}

\vspace*{1.5ex}

{\large \it Harish-Chandra Research Institute\footnote{Formerly
Mehta Research Institute of Mathematics \&\ Mathematical Physics}}\\
{\large \it  Chhatnag Road, Jhusi, Allahabad 211019, India}

\vspace*{4.5ex}

{\bf Abstract}

\begin{quote}

We show that the world-volume theory on a D-p-brane at the tachyonic
vacuum has solitonic string solutions whose dynamics is governed by the
Nambu-Goto action of a string moving in (25+1) dimensional space-time. 
This provides strong evidence for the conjecture that at this vacuum the
full (25+1) dimensional Poincare invariance is restored. We also use this
result to argue that the open string field theory at the tachyonic vacuum
must contain closed string excitations. 

\end{quote}
\end{center}

\newpage

\tableofcontents

\baselineskip=16.9pt

\sectiono{Introduction and Summary} \label{s1}

It has been conjectured that the tachyonic vacuum in open bosonic string
theory on a D-brane describes the closed string vacuum without D-branes,
and that various soliton solutions in this theory describe D-branes of
lower dimension\cite{9902105}. Similar conjectures have also been put
forward for superstring theories\cite{ORIGINAL,9810188,9812135}. Evidence
for these conjectures come from both, first\cite{RECK,FIRST} and
second\cite{SECOND,0005031,NONCOM,BSFT} quantised string theories. 

Given that all D-branes can be regarded as solitons in the open string
field theory\cite{SFT}, one might wonder if the open string field theory
could be used for a non-perturbative formulation of string
theory\cite{9904207}. For this one needs to show that not only the
D-branes, but other known objects in string theory, namely the fundamental
closed strings and the NS five-branes are also present in this open string
field theory.  Progress in identifying the fundamental string has been
made in refs.\cite{9901159,0002223,0005031,0009061,0010181}. In
particular, in \cite{0002223,0005031,0009061,0010181} it was shown that
the effective action\cite{9909062} describing the dynamics of the D-brane
around the tachyonic vacuum admits string-like classical solution whose
tension matches that of a fundamental string. It was also established that
on a D-$25$-brane world-volume, the dynamics of these strings is described
by that of a Nambu-Goto string moving in $(25+1)$ dimensions. 

On the world-volume of a D-$p$-brane embedded in the (25+1) dimensional
space-time, the full (25+1) dimensional Poincare invariance is
spontaneously broken to the product of $(p+1)$ dimensional Poincare group,
and the $(25-p)$ dimensional rotation group. However, if the tachyonic
ground state really represents the vacuum without a D-brane, then we
expect that in this vacuum the full (25+1) dimensional Poincare invariance
should be restored. Thus the dynamics of the string-like solutions should
be described by a Nambu-Goto action with (25+1) dimensional target space
rather than a $(p+1)$ dimensional target space. This is what we shall
demonstrate in this paper.\footnote{This question was partially addressed
in \cite{0009061} where it was shown that in the approximation where the
contribution to the Hamiltonian is dominated by the electric flux on the
D-brane world-volume, there is a symmetry that exchanges the velocity
tangential to the D-brane with the velocity transverse to the D-brane.
} Since the Nambu-Goto action in (25+1)
dimensional target space has full
(25+1)  dimensional Poincare invariance, this result provides a strong
support to the conjecture that at the tachyonic vacuum of the D-$p$ brane
the full (25+1) dimensional Poincare invariance is
restored.\footnote{Earlier string field theory analysis has provided
evidence for the restoration of the translational invariance along
directions transverse to the D-brane world-volume\cite{0007153,0008033}.}

Since the D-$p$-brane world-volume is $(p+1)$ dimensional, and the string
solution lives on the D-$p$-brane, it may sound strange at first that this
string actually moves in (25+1) dimensions. The reason it can happen is
that at the tachyonic vacuum the D-brane has vanishing tension, and hence
it does not cost the D-brane any additional energy to adjust its
world-volume to contain any given fundamental string world-sheet embedded
in (25+1) dimensional space-time. Thus the string world-sheet
always lies inside the D-$p$-brane world-volume, as should be the case.
The non-trivial fact here is that the dynamics of the string
tangential to the D-brane world-volume, which is described by
the gauge fields, and the dynamics transverse to
the D-brane world-volume, which is described by the massless scalar fields
associated with the transverse motion of the D-brane, are together
described by the Nambu-Goto action in the full (25+1) dimensional target
space-time.

Although the dynamics of the string soliton constructed this way agrees
with that of the fundamental string, there are some caveats. First of all
the tension of the string is governed by the total amount of electric flux
it carries, and only after properly taking into account the quantization
rule for the electric flux one can show that the tension matches that of
the fundamental string. Within the classical field theory which we shall
be studying, there is no rationale for this quantization law. A related
problem is as follows. Although the string solution constructed here has
the correct degrees of freedom describing the dynamics of a fundamental
string, it also has additional degrees of freedom corresponding to the
energy density spreading out in the direction transverse to the original
string
solution instead of being confined in a thin tube along the string.
We show that these problems can be avoided by making the solitonic string
driven by an external open string. For this we consider the case
where one of the directions transverse to the D-$p$-brane is compact, and
we begin with a configuration of open strings starting on the D-brane, and
ending on its image under translation along the compact direction. We
then ask what happens when the tachyon on the D-brane rolls down to its
ground state. We argue that at the tachyonic vacuum, the two ends of the
original open string are connected by a flux line on the D-brane, with the
total amount
of flux fixed by the source (and the sink) of flux, namely the end points
of the original open string on the brane. Furthermore, the condition
for minimum energy prevents the flux from spreading, since the source and
the sink of flux are
point-like objects on the D-$p$-brane world-volume. The net result is a
single fundamental string winding along the compact direction. Using a
T-duality
transformation along the compact direction we can then argue that the
T-dual D-$(p+1)$-brane at the tachyonic vacuum must contain closed
string excitations carrying momentum along the compact direction.

Related earlier work in refs.\cite{LIND} analysed the dynamics of
tensionless D-branes in a different formalism and found that the D-brane
world-volume is foliated by string world-sheet. It will be interesting to
explore the precise relation between these results and the
static gauge results of
refs.\cite{0002223,0005031,0009061,0010181} and the
present paper. 

The paper is organised as follows. In section \ref{s2} we review the
result for the effective action on the D-brane world-volume at the
tachyonic vacuum\cite{9909062} and its Hamiltonian
formulation\cite{0009061}. In
section \ref{s3} we show that given any solution of the equations of
motion of a Nambu-Goto string moving in (25+1) dimensional space-time, we
can construct a solution of the equations of motion of the D-$p$-brane
world-volume theory with energy density localised along the world-sheet of
the corresponding Nambu-Goto string solution. This establishes that the
D-$p$-brane world-volume theory admits string-like soliton solutions whose
dynamics is governed by the Nambu-Goto action in (25+1) dimensions. In
section \ref{s4} we use this result to argue that the open string field
theory, describing the D-brane world-volume theory at the tachyonic
vacuum, must contain closed string excitations.

\sectiono{Low Energy Effective Field Theory on the D-brane at the
Tachyonic Vacuum} \label{s2}

We shall analyse the dynamics of massless fields living on a D-$p$ brane
at the tachyonic vacuum in the static gauge. Let us denote by $x^\mu$
($0\le \mu\le
p$) the world-volume coordinates on the D-brane, by $A_\mu$  the
U(1) gauge field living on the D-brane, and by $Y^I$ ($p+1\le I\le 25$) 
the massless scalars
representing the transverse coordinates of the brane. 
The action is given by\cite{9909062}:
\be \label{eaction}
S = - V(T) \int d^{p+1} x \sqrt{-\det(\eta_{\mu\nu} + F_{\mu\nu} + \p_\mu
Y^I
\p_\nu Y^I)}\, ,
\ee
where $V(T)$ is the tachyon potential which vanishes at the tachyonic
vacuum $T=T_0$.
We shall work in the
gauge $A_0=0$, and denote by $\pi^i(x)$ and $p_I(x)$ the momenta conjugate
to $A_i$ and $Y^I$ respectively for $1\le i\le p$. As was shown in
\cite{0009061}, the dynamics of the brane at the tachyonic vacuum is best
described in the Hamiltonian formalism, with the Hamiltonian
\be \label{e2.1}
H = \int d^p x \HH\, ,
\ee
with 
\be \label{e2.2}
\HH = \sqrt{\pi^i\pi^i + p_I p_I + (\pi^i \p_i Y^I)^2 + b_i b_i}\, ,
\ee
where
\be \label{e2.3}
b_i \equiv F_{ij} \pi^j + \p_i Y^I p_I\, .
\ee
$F_{ij}=\p_i A_j - \p_j A_i$ is the magnetic field strength. The $\pi^i$'s
satisfy a constraint:
\be \label{e2.4}
\p_i\pi^i = 0\, .
\ee
In writing down the Hamiltonian \refb{e2.1}, \refb{e2.2} we have taken the
tachyon field $T$ to be frozen at its minimum $T=T_0$.\footnote{Proposals
for the effective action including tachyon kinetic term have been put
forward in \cite{TACHYONIC}.}

Let us denote by $E_i=\p_0 A_i$ the electric field strength. Then the
Bianchi identities and the
equations of motion derived from the Hamiltonian given in \refb{e2.1},
\refb{e2.2} are given by
\be \label{e2.5}
\p_{[i} F_{jk]} = 0, \qquad \p_0 F_{ij} = \p_i E_j - \p_j E_i\, ,
\ee
\be \label{e2.6}
E_i = {1\over \HH} (\pi^i + \p_i Y^I \pi^j \p_j Y^I - F_{ij} b_j)\, ,
\ee
\be \label{e2.7}
\p_0 \pi^i + \p_j \Big({1\over \HH} (\pi_j b_i - \pi_i b_j)\Big) = 0\, ,
\ee
\be \label{e2.8}
\p_0 Y^I = {1\over \HH} (p_I + \p_i Y^I b_i)\, ,
\ee
\be \label{e2.9}
\p_0 p^I = \p_i\Big( {1\over \HH} (\pi^i \pi^j \p_j Y^I + b_i p^I) 
\Big)\, .
\ee

For this system, there are conserved Noether currents $T_{\mu\nu}$
and $T_{\mu I}$ ($0\le\mu,\nu\le p$, $(p+1)\le I\le 25$) associated
with the translation along the spatial coordinates $x^\mu$ labelling the
D-$p$-brane world-volume, as
well as translation along
the coordinates $Y^I$ transverse to the world-volume. These are given by: 
\ben \label{e2.10}
&& T_{00} = \HH, \qquad T_{k0} = - b_k\, ,
\qquad T_{0i} = - b_i, \qquad T_{ki} = {1\over \HH}
(\pi^k \pi^i - b_k b_i)\, , \nonumber \\
&& T_{0I} = p_I, \qquad T_{kI} = {1\over \HH} (\pi^k \pi^j \p_j Y^I + b_k
p^I)\, ,
\een
and satisfy
\be \label{e2.11} 
\eta^{\mu\nu} \p_\mu T_{\nu\rho} = 0, \qquad \eta^{\mu\nu} \p_\mu T_{\nu
I} = 0\, .
\ee

\sectiono{Fundamental String Solution} \label{s3}

In this section we shall demonstrate that the equations of motion
discussed in section \ref{s2} admit fundamental string solutions whose
dynamics is identical to that of a Nambu-Goto string moving in (25+1)
dimensional space-time. Using the results of \cite{NO}, 
ref.\cite{0009061} showed that if
we set the $Y^I$'s to 0, then the dynamics of the solitonic string is
described by the Nambu-Goto action in
$(p+1)$ dimensional space-time. The
new result is the incorporation of the $Y^I$'s. Since
the dynamics of a Nambu-Goto string in (25+1) dimensional space-time has
full (25+1) dimensional Poincare invariance, our result gives strong
support to the
conjecture that the tachyonic vacuum of the D-$p$-brane represents a
configuration where the full (25+1) dimensional Poincare invariance is
restored. 

Our strategy will be as follows. We shall show that for every
configuration of a Nambu-Goto string satisfying the string equations of
motion we can construct a solution of the equations of motion
\refb{e2.4} - \refb{e2.9}, with energy density localised along the string.
For this we start by writing down the action of the Nambu-Goto string in
(25+1) dimensional space-time:
\be \label{e3.1}
S_{NG} = -\int d\tau \, d\sigma \, \sqrt{-\det(\eta_{MN} \p_\alpha Z^M
\p_\beta Z^N)}
\ee
where $\xi^\alpha$ for $\alpha=0,1$ denote the world-volume coordinates
of the string:
$(\xi^0,\xi^1)\equiv (\tau,\sigma)$, $Z^M$ ($0\le M\le 25$) denote the
space-times coordinates of the string, and $\eta_{MN}$ is the
Minkowski metric $diag(-1,1,1,\ldots 1)$. We shall choose the static
gauge: $(Z^0=\tau, Z^1=\sigma)$ and go to the Hamiltonian formalism. If we
denote by $P_s$ the momenta conjugate to $Z^s$ for $2\le s\le 25$, the
Hamiltonian is given by:
\be \label{e3.2}
H_{NG} \equiv \int d\sigma \, \HH_{NG} = \int d\sigma \, \sqrt{1 + P_s P_s
+ \p_\sigma Z^s \p_\sigma Z^s + (P_s \p_\sigma Z^s)^2}\, .
\ee
The equations of motion following from this Hamiltonian is given by:
\be \label{e3.3}
\p_\tau Z^s = {1\over \HH_{NG}} (P_s + \p_\sigma Z^s P_t \p_\sigma Z^t)\,
,
\ee
\be \label{e3.4}
\p_\tau P_s = \p_\sigma\Big( {1\over \HH_{NG}} (\p_\sigma Z^s + P_s P_t
\p_\sigma Z^t)\Big)\, .
\ee
In these equations $s$ and $t$ indices take values
$2,3,\ldots 25$. For future use, we shall define
\be \label{e3.5}
P_1 = - \sum_{s=2}^{25} P_s \p_\sigma Z^s\, , \qquad Z^1(\tau,\sigma) =
\sigma\, .
\ee
With these definitions, it is straightforward to verify that
eqs.\refb{e3.3}, \refb{e3.4} are satisfied also for $s=1$. (The sum over
$t$ in these equations still runs from 2 to 25).

Let $(Z^s(\tau,\sigma), P_s(\tau,\sigma))$ for $2\le s\le 25$ be a
solution of eqs.\refb{e3.3}, \refb{e3.4}. Now consider the following field
configuration on the D-$p$-brane:
\ben \label{e3.6} 
\pi^i (x^0,\ldots x^p) &=& \p_\sigma Z^i (\tau, \sigma)
f(x^0,\ldots x^p)\Big|_{(\tau,\sigma)=(x^0, x^1)},
\nonumber
\\
p_I(x^0, \ldots x^p) &=& P_I(\tau,\sigma) f(x^0,\ldots x^p)
\Big|_{(\tau,\sigma)=(x^0, x^1)} \nonumber \\
\een
where we have used the convention that the indices $i,j,k$ run from 1 to
$p$, the indices $I,J,K$ run from $(p+1)$ to 25, and the indices $s,t$ run
from 2 to 25.
$f(x^0,\ldots x^p)$ is an arbitrary function of the variables $(x^m-
Z^m(x^0,x^1))$ for $2\le m\le p$, and hence satisfies:
\be \label{e3.6a}
\p_\sigma Z^i \p_i f\Big|_{(\tau,\sigma)=(x^0, x^1)}=0,
\qquad (\p_\tau Z^i \p_i f + \p_0 f)\Big|_{(\tau,\sigma)=(x^0, x^1)}=0.
\ee
The fields $Y^I(x^0,\ldots x^p)$ and $F_{ij}(x^0,\ldots
x^p)$ are subject to the following set of conditions:
\be \label{e3.6ca}
(\p_\sigma Z^j \p_j Y^I - \p_\sigma Z^I)\Big|_{(\tau,\sigma)=(x^0,
x^1)} = 0\, , 
\ee
and 
\be \label{e3.6cb}
(F_{ij} \p_\sigma Z^j + \p_i Y^I P_I + P_i)\Big|_{(\tau,\sigma)=(x^0,
x^1)} = 0\, ,
\ee
but are otherwise unspecified.
Using eqs.\refb{e3.6}, \refb{e3.6ca} and \refb{e3.6cb} we can easily
verify that for this
background,
\ben \label{e3.6d}
\HH(x^0,\ldots x^p) &=& H_{NG} (\tau=x^0, \sigma=x^1) f(x^0,\ldots x^p)\,
, \nonumber \\
b_i(x^0,\ldots x^p) &=& - P_i (\tau=x^0, \sigma=x^1) f(x^0,\ldots
x^p)\, , \nonumber \\
\pi^j \p_j Y^I (x^0, \ldots x^p) &=& \p_\sigma Z^I 
(\tau=x^0, \sigma=x^1) f(x^0,\ldots
x^p)\, .
\een
Using eqs.\refb{e3.3}-\refb{e3.6a} and
\refb{e3.6d} we can now verify
that eqs.\refb{e2.4}, \refb{e2.7} and \refb{e2.9} are satisfied by this
background. Thus in order to construct a solution of the full
set of equations of motion \refb{e2.4}-\refb{e2.9} we need to show that it
is possible to find $F_{\mu\nu}$ and $Y^I$ satisfying the constraints
\refb{e2.5}, \refb{e2.6}, \refb{e2.8}, \refb{e3.6ca} and \refb{e3.6cb}.

First we shall establish the existence of $Y^I$'s satisfying
eqs.\refb{e2.8} and \refb{e3.6ca}. (Note that the eq.\refb{e3.6cb} imposes
a constraint on $Y^I$ of the form $\p_\sigma Z^i(\p_i Y^I P_I + P_i)
\Big|_{(\tau,\sigma)=(x^0,
x^1)} = 0$, but due to eq.\refb{e3.5}, this is automatically satisfied
once eq.\refb{e3.6ca} is
satisfied.) Using
eqs.\refb{e3.6}, \refb{e3.6d}, we shall now
write eqs.\refb{e2.8} and \refb{e3.6ca} as follows:
\ben \label{e3.n1}
\p_0 Y^I &=& {1\over \HH_{NG}} (P_I - \p_i Y^I P_i)\, ,\nonumber \\
\p_1 Y^I &=& (- \p_\sigma Z^m \p_m Y^I + \p_\sigma
Z^I)\, ,
\een
where the indices $m,n,q$ run from 2 to $p$, and it will be understood
from now on that
$\tau$ and $\sigma$ are to be identified with $x^0$ and $x^1$
respectively. We can now treat eqs.\refb{e3.n1} as the equations
determining the $x^0$ and $x^1$ evolution of the functions $Y^I$. (We
replace the $\p_1 Y^I$ appearing on the right hand side of the first
equation by the right hand side of the second equation.)
Existence of a solution to these equations requires the integrability
condition:
\be \label{e3.n2}
\p_1 \Big( {1\over \HH_{NG}} \big(P_I - \p_m Y^I P_m
+ P_1 (\p_\sigma Z^m \p_m Y^I - \p_\sigma Z^I)\big) \Big)
- \p_0 \Big( (- \p_\sigma Z^m \p_m Y^I + \p_\sigma
Z^I) \Big) = 0\, .
\ee
It is a straightforward although tedious exercise to show that once
eqs.\refb{e3.3}, \refb{e3.4} are satisfied, eq.\refb{e3.n2} is satisfied.

Thus it remains to show the existence of a set of $F_{\mu\nu}$ satisfying
eqs.\refb{e2.5}, \refb{e2.6} and \refb{e3.6cb}. We begin with the
$F_{mn}$'s ($2\le m,n,q \le p$). We take them to satisfy the following
identities:
\be \label{e3.n3}
\p_{[m}F_{nq]} = 0\, ,
\ee
and 
\ben \label{e3.n4}
&& \p_1 F_{mn} + \p_1 Z^q[x^0,x^1] \p_q F_{mn} = 0\, , \nonumber \\
&& \p_0 F_{mn} + \p_0 Z^q[x^0, x^1] \p_q F_{mn} = 0\, .
\een
To see that it is possible to choose $F_{mn}$'s satisfying these
conditions, we regard eqs.\refb{e3.n4} as the evolution equation for
$F_{mn}$ in $x^0$ and $x^1$ from an initial configuration satisfying the
Bianchi identities
\refb{e3.n3}. It is easy to verify that the evolution equations
\refb{e3.n4} preserve the Bianchi identities at all values of $x^0$ and
$x^1$. It is also easy to verify the integrability of the equations
\refb{e3.n4}:
\be \label{enn1}
\p_0 \big(\p_1 Z^q[x^0,x^1] \p_q F_{mn}\big) - \p_1\big(\p_0 Z^q[x^0, x^1]
\p_q F_{mn} \big) = 0\, .
\ee

Given $F_{mn}$ satisfying \refb{e3.n3}, \refb{e3.n4}, we can use
\refb{e3.6d} to write
eqs.\refb{e2.6} and
\refb{e3.6cb} as follows:
\be \label{e3.n5}
F_{0i} = {1\over \HH_{NG}} (\p_\sigma Z^i + \p_i Y^I \p_\sigma Z^I
+ F_{ij} P_j)\, ,
\ee
and 
\be \label{e3.n6}
F_{i1} = - F_{in} \p_\sigma Z^n - \p_i Y^I P_I  - P_i\, .
\ee
If eq.\refb{e3.n6} is satisfied for $i=m$, then it is also automatically
satisfied for $i=1$ with the help of eqs.\refb{e3.5} and \refb{e3.n1}.
Thus the independent equations in \refb{e3.n6} are:
\be \label{e3.n7}
F_{m1} = - F_{mn} \p_\sigma Z^n - \p_m Y^I P_I  - P_m\, .
\ee
This gives $F_{m1}$ in terms of $F_{mn}$ and other known quantities.
Replacing the $F_{m1}$'s appearing on the right hand side of
eq.\refb{e3.n5} by
the right hand side of eq.\refb{e3.n7}, we can now regard eqs.\refb{e3.n5}
as
expressions
for $F_{01}$ and $F_{0m}$ in terms of $F_{mn}$ and other known quantities.

We now need to check that $F_{0i}$ and $F_{m1}$ defined through
eqs.\refb{e3.n5}, \refb{e3.n7} satisfy the remaining Bianchi identities:
\ben \label{e3.n8}
&& \p_0 F_{mn} + \p_m F_{n0} + \p_n F_{0m} = 0\, , \nonumber \\
&& \p_1 F_{mn} + \p_m F_{n1} + \p_n F_{1m} = 0\, , \nonumber \\
&& \p_0 F_{m1} + \p_m F_{10} + \p_1 F_{0m} = 0\, .
\een
It is straightforward to verify that
all of these identities are consequences of eqs.\refb{e3.3}, \refb{e3.4},
\refb{e3.n1}, \refb{e3.n3} and \refb{e3.n4}. This completes the
construction of a solution of the complete set of equations of motion
\refb{e2.4}-\refb{e2.9} of the D-$p$-brane world-volume field theory.

We shall now make a special choice of the function $f$:
\be \label{e3.6b}
f(x^0,\ldots x^p) = \prod_{m=2}^p
\delta(x^m - Z^m(x^0, x^1)) \, ,
\ee
which satisfies eq.\refb{e3.6a}. Furthermore we take
\be \label{e3.n9}
Y^I(x^0, x^1, x^m=Z^m(x^0, x^1)) = Z^I(x^0, x^1)\, ,
\ee
which can be seen to be compatible with eqs.\refb{e3.n1}.\footnote{Indeed,
eqs.\refb{e3.n1} and \refb{e3.n4} can be interpreted as the requirement of
vanishing of the derivatives of $(Y^I-Z^I)$ and $F_{mn}$ along directions
tangential to the string world-sheet. Thus we can solve these equations by 
taking $Y^I-Z^I$ and
$F_{mn}$ to be arbitrary functions $g^I$ and $g_{mn}$  respectively of
$x^2 - Z^2(x^0,
x^1), \ldots x^p -
Z^p(x^0, x^1)$. Eq.\refb{e3.n9} can then be satisfied by requiring that
$g^I(0,\ldots 0)=0$. The Bianchi identities \refb{e3.n3} are
satisfied by
requiring that the functions $g_{mn}$ satisfy $\p_{[q}g_{mn]}=0$.} As
can be
seen from eqs.\refb{e3.6d}, for the choice of $f$ given
in eq.\refb{e3.6b}, the energy density is localised along the surface
$x^m=Z^m(x^0, x^1)$ for $2\le m\le p$. Using eq.\refb{e3.n9} we see that
in the full (25+1) dimensional space-time, this describes the surface $x^s
=
Z^s(x^0, x^1)$ for $2\le s\le 25$. This is precisely the world-sheet of
the
string described by the Nambu-Goto action \refb{e3.1}. Thus our analysis
shows that whenever the Nambu-Goto equation has
a solution described by $Z^s(\sigma,\tau)$, there is a corresponding
solution in the D-$p$-brane world-volume field theory with energy density
localised along the world-sheet of the string. In other words, the
D-$p$-brane world-volume theory contains a solution whose dynamics is
exactly that of a Nambu-Goto string in (25+1)-dimensions.

Note, however, that the freedom of replacing the $\delta$-function by an
arbitrary function of $x^m - Z^m(x^0,x^1)$ means that besides the usual
degrees of freedom of the fundamental string, our solution has additional
degrees of freedom which corresponds to the freedom of spreading out the
electric flux in directions transverse to the string. Also the overall
normalization on the right hand side of eq.\refb{e3.6b}, which fixes the
tension / charge of the string, is arbitrary. We
shall return to these questions in the next section. There are also
additional degrees of freedom stemming from the fact that
eqs.\refb{e3.n1},
\refb{e3.n3} and \refb{e3.n4} do not determine $Y^I$ and $F_{mn}$
completely for a given configuration of the Nambu-Goto string. This is
analogous to the spurious degeneracy found in \cite{SPURIOUS}. It
has been argued in \cite{AMBIG} that this apparent degeneracy is due to
the wrong choice of variables in describing the theory, and will disappear
once we use the correct set of variables.

We shall end this section by writing down the expressions for the
conserved Noether currents for the background described above. This is a
straightforward exercise, and the results are as follows:
\ben \label{e3.8}
&& T_{00}(x^0,\ldots x^p) = \HH_{NG}(\tau,\sigma)\prod_{m=2}^p
\delta(x^m - Z^m(\tau, \sigma)) \Big|_{(\tau,\sigma)=(x^0, x^1)}, \nonumber
\\
&& T_{0k}(x^0,\ldots x^p) = T_{k0}(x^0,\ldots x^p) 
= P_k(\tau,\sigma) \prod_{m=2}^p
\delta(x^m - Z^m(\tau, \sigma)) \Big|_{(\tau,\sigma)=(x^0, x^1)}, \nonumber
\\
&& T_{ki}(x^0,\ldots x^p) = {1\over \HH_{NG}} 
(\p_\sigma Z^k\p_\sigma Z^i - P_k P_i)
\prod_{m=2}^p
\delta(x^m - Z^m(\tau, \sigma)) \Big|_{(\tau,\sigma)=(x^0, x^1)}, \nonumber
\\
&& T_{0I}(x^0,\ldots x^p) = P_I(\tau,\sigma) \prod_{m=2}^p
\delta(x^m - Z^m(\tau, \sigma)) \Big|_{(\tau,\sigma)=(x^0, x^1)}, \nonumber
\\
&& T_{kI}(x^0,\ldots x^p) = {1\over \HH_{NG}} 
(\p_\sigma Z^k \p_\sigma Z^I - P_k P_I)
\prod_{m=2}^p
\delta(x^m - Z^m(\tau, \sigma)) \Big|_{(\tau,\sigma)=(x^0, x^1)}. 
\nonumber
\\
\een
Verification of the conservation laws \refb{e2.11} for $T_{\mu \nu}$ and
$T_{\mu I}$ is a straightforward application of eqs.\refb{e3.3},
\refb{e3.4}. It is also a simple exercise to verify that the corresponding
conserved charges $\int d^p x\, T_{00}$, $\int d^p x T_{0i}$ and $\int d^p
x\, T_{0I}$ agree with the Noether charges of the Nambu-Goto string
associated with translation invariance along $x^0$, $x^i$ and $x^I$
directions respectively.

\sectiono{Closed Strings in the D-brane World-Volume Theory} \label{s4}

In this section we shall use the results of the previous section to argue
that at the tachyonic vacuum the D-brane world-volume theory must contain
closed string excitations. 
Identification of closed strings as closed flux lines
has been discussed earlier in
refs.\cite{9901159,0002223,0005031,0009061,0010181}. The present
construction is closely related, but differs in that here part of the
closed string is formed by an external open string. 

We begin with a thought experiment. Consider
three well separated D-branes $A$, $B$ and $C$, and a state on the
world-volume of this system consisting of a fundamental string stretched
from $A$ to $B$, and another fundamental string stretched from $B$ to $C$. 
Let us now ask: what happens to this state when the tachyon field on the
brane $B$ rolls down to its (local) minimum, but the branes $A$ and $C$
remain unchanged. 
The D-brane world-volume field
theory analysis tells us that the ends of the $AB$ and $BC$ strings are
source and
sink of one unit of electric flux (measured in natural units) on the
world-volume of the brane $B$. Thus the
fate of the system is clear: the final configuration will consist of the
$AB$ string, the $BC$ string, and an electric flux line (described by
the solution given in section \ref{s3}) on the $B$-brane
world-volume connecting the end point of the $AB$ string to the starting
point of the $BC$ string. Note that the condition
for minimum energy
will prevent the flux from spreading
out as its source and sink are point objects.\footnote{Of course, one
would still need to understand why local fluctuations on the string
involving the spreading of the flux is absent. Some discussion on this can
be found in refs.\cite{0002223,0005031}.} Furthermore there is
precisely one unit of electric flux and hence its tension is equal to that
of a fundamental string\cite{0002223}. Thus it is natural to interprete
the flux
line as a fundamental string stretched from the end point of the $AB$
string to the starting point of the $BC$ string. (This string, as
well as the external $AB$ and $BC$ strings, can
adjust their positions and orientations so as to minimise the energy of
the configuration).
Thus the net result of this process is a single open string stretched
between $A$ and $C$. It is as if the tachyon condensation on the
world-volume of the brane $B$ joins the ends of the $AB$ and $BC$
strings by a fundamental string. Even before tachyon condensation
on the brane $B$, the
ends of $AB$ and $BC$
strings could join to produce an $AC$ string. But there it was a
perturbative quantum process, whereas the process described here is a
non-perturbative classical process.

Now we consider a different system: take a single D-brane and an
open string with both ends on this D-brane. One can use the same argument
to conclude that when the tachyon condenses to its ground state, the two
ends of the open string will be connected by a flux line, and once we
identify the flux line as the fundamental string, we get a closed
string state! This argument can be made more concrete as follows. Take
a D-brane with one of
its transverse directions compact, and consider an open string stretched
from the D-brane to its image under translation along the compact
direction. Now let us ask what happens to this open string state when the
tachyon on the D-brane condenses to its ground state. Since the original
state carries fundamental string winding charge this state cannot
disappear. To see
what happens it is simplest to go to the
infinite cover; in this case we have initially an infinite number of
parallel
D-branes at regular spacing, and between any two neighbouring D-branes we
have an open string suspended between the two. Thus on any of the D-branes
we have an open string ending and another open string starting giving a
source and a sink of electric flux. When the
tachyon condenses to its ground state, each D-brane develops a flux line
joining the source to the sink. If we identify this flux line as a
fundamental string as before, the result is a single infinitely long
string. After modding out by the discrete
translation
symmetry to compactify the direction, this is nothing but a closed string
wrapped around the compact direction!

Thus we conclude that if we start with a D-brane with a transverse
direction compact, then after tachyon condensation the open string field
theory on the D-brane world-volume must contain excitations corresponding
to closed string winding states along the compact direction. Let us now
make a T-duality transformation along the compact direction. This converts
the D-$p$ brane to a D-$(p+1)$ brane, but is otherwise a symmetry of the
open string field theory order by order in open string perturbation
theory. On the other hand this transforms the closed string winding modes
to closed string momentum modes. 
Thus if we start with a D-brane with one of its
tangential directions compact, then after tachyon condensation the
corresponding open
string field theory will contain excitations corresponding to closed
string states carrying momentum along the compact direction. 

It will be
interesting to see if we can study this phenomenon directly in open string
field theory.

\noindent{\bf Acknowledgement}: I would  like to thank S. Das, D. Ghoshal,
D. Jatkar, J. Majumder and P. Mukhopadhyay  for useful
discussions.

\end{document}